\documentclass[sigplan,screen]{acmart}
\usepackage{amsthm}
\usepackage{amsmath}
\usepackage{algorithm}
\usepackage{algpseudocode}

\AtBeginDocument{%
  }

\theoremstyle{remark}

\newtheorem{theorem}{Theorem}
\newtheorem{lemma}[theorem]{Lemma}

\begin{document}

\title{Operational Safety in Human-in-the-loop Human-in-the-plant Autonomous Systems}

 \author{Ayan Banerjee$^1$, Aranyak Maity$^1$, Imane Lamrani$^2$, Sandeep K.S. Gupta$^1$}
\email{abanerj3@asu.edu, amaity1@asu.edu, ilamrani@asu.edu, sandeep.gupta@asu.edu}
 \affiliation{%
   \institution{$^1$School of Computing and Augmented Intelligence, Arizona State University, $^2$ Nikola Motors}
  \city{Tempe,Az}
   \country{USA}
 }

\newcommand{\indicator}[1]{\mathbf{1}_{\{#1\}}}
\begin{abstract}
  Control affine assumptions, human inputs are external disturbances, in certified safe controller synthesis approaches are frequently violated in operational deployment under causal human actions. This paper takes a human-in-the-loop human-in-the-plant (HIL-HIP) approach towards ensuring operational safety of safety critical autonomous systems: human and real world controller (RWC) are modeled as a unified system. A three-way interaction is considered: a) through personalized inputs and biological feedback processes between HIP and HIL, b) through sensors and actuators between RWC and HIP, and c) through personalized configuration changes and data feedback between HIL and RWC. We extend control Lyapunov theory by generating barrier function (CLBF) under human action plans, model the HIL as a combination of Markov Chain for spontaneous events and Fuzzy inference system for event responses, the RWC as a black box, and integrate the HIL-HIP model with neural architectures that can learn CLBF certificates. We show that synthesized HIL-HIP controller for automated insulin delivery in Type 1 Diabetes is the only controller to meet safety requirements for human action inputs.
\end{abstract}

\maketitle

\section{Introduction}

Safety criticality implies that the operation of the autonomous system (AS) can cause harm to the human participants who are affected by the AS goal~\cite{Banerjee15TMCV2,Lamrani21OperationalV2,lamrani2018hymn}. Given the impending risks to the human user, safety critical applications most typically operate with a human in the loop (HIL). The human is in charge of starting and stopping automation and can provide manual inputs whenever the user perceives safety risks or operational inefficiency. In medical applications such as automated insulin delivery, this results in a HIL-human in the plant (HIL-HIP) system model (Section \ref{sec:sysmod})~\cite{Banerjee23High,banerjee2024cpsllmlargelanguagemodel}, where the human user is the monitor / decision maker and also part of the physical plant controlled by the AS (Fig. \ref{fig:newSysMod}). 

Existing safety certification process~\cite{banerjee2011ensuring,banerjee2013using,bagade2017validation,waise}, assume a control affine system model, where the plant state $X$ is assumed to follow the dynamics in Eqn \ref{eqn:Plant}.
\begin{equation}
\label{eqn:Plant}
\scriptsize
\dot{X} = f(X)+g(X)\pi(X),
\end{equation}
where $f(.)$ is the un-perturbed plant response model, $g(.)$ is the input effect, and $\pi(.)$ is a controller that computes an input to the plant based on the plant state $X$. In a HIL-HIP architecture, the input to the plant is given by: $u=\pi(X)+u_{ex}$, where $u_{ex}\in U_{ex}$ is an external input from the human user. Despite human user being an integral part of AS operation, safety assurance using control affine assumption consider human as external to the system. As such an ``average user" is considered under specific operational scenarios so that human inputs $U_{ex}$ is modeled as a noise disturbance with known probability distribution.

Large scale deployment and day-to-day usage imply that a significant number of users will be non-conformal to the ``average user" resulting in novel usage scenarios. To replicate the performance obtained in the certification process, the real user may undertake \textit{personalization actions}, which are external inputs or system configuration changes applied with/without expert advisory agent (clinicians) consultations. Such inputs may have a causal relation with the HIP state $X$, are out of distribution, and may invalidate safety certificates with control affine assumption. Such unverified personalizations can jeopardize operational safety~\cite{banerjee2012-your-mobility,banerjee2024cpsllmlargelanguagemodel}. 

\begin{figure}
\includegraphics[width=\columnwidth,trim = 0 20 0 0]{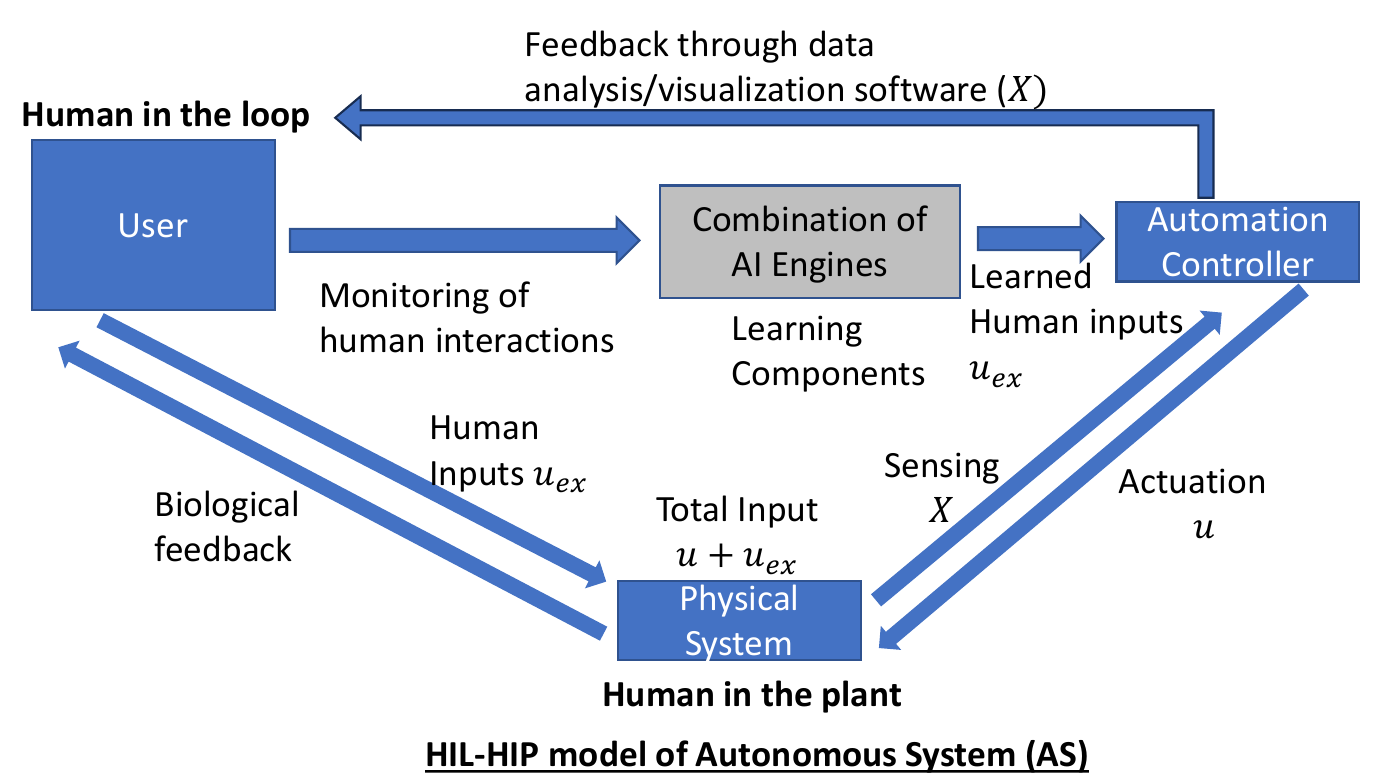}
\caption{HIL-HIP autonomous systems (AS).}
\label{fig:HIL-HIP}
\end{figure}

In this paper, we assume that the real world controller (RWC) or $\pi(.)$ in AS is \underline{safety certified} with control affine assumption, for the ``average user" and a \underline{black box}, and present a technique to derive safety certificates of AS with HIL-HIP architecture where the plant model is given by:
\begin{equation}
\label{eqn:FG}
\scriptsize
\dot{X} = f(X)+g(X)(\pi(X)+u_{ex}),    
\end{equation}
 and $u_{ex} \in U_{ex}$ is a set of personalized inputs for a given real life user. We validate the presented technique by synthesizing a certified safe controller for automated insulin delivery (AID) to control glucose levels in Type 1 Diabetes (T1D).
\section{System Model and preliminaries}
\label{sec:sysmod}
A \textbf{plant} is described by the $N$ dimensional real state vector $X\in \mathcal{X}$, where $\mathcal{X} \subset \mathcal{R}^N$ is the state space of the plant. For AID, $X$ is a $3 \times 1$ vector, with CGM glucose, interstitial insulin concentration, and plasma insulin concentration as elements. 

\noindent An \textbf{autonomous real world controller} (RWC) $\pi(X)$ uses sensors (CGM for AID) on the plant (human body) to monitor its current state, and actuators (insulin pump) to deliver control inputs $u \in \mathcal{R}$ (micro bolus insulin). The control task of the RWC is to drive a state variable $x_i \in X$ (CGM) to a set point $x^g_i$ (typically 120 mg/dL). 

\begin{figure}
\includegraphics[width=\columnwidth,trim=0 50 0 0]{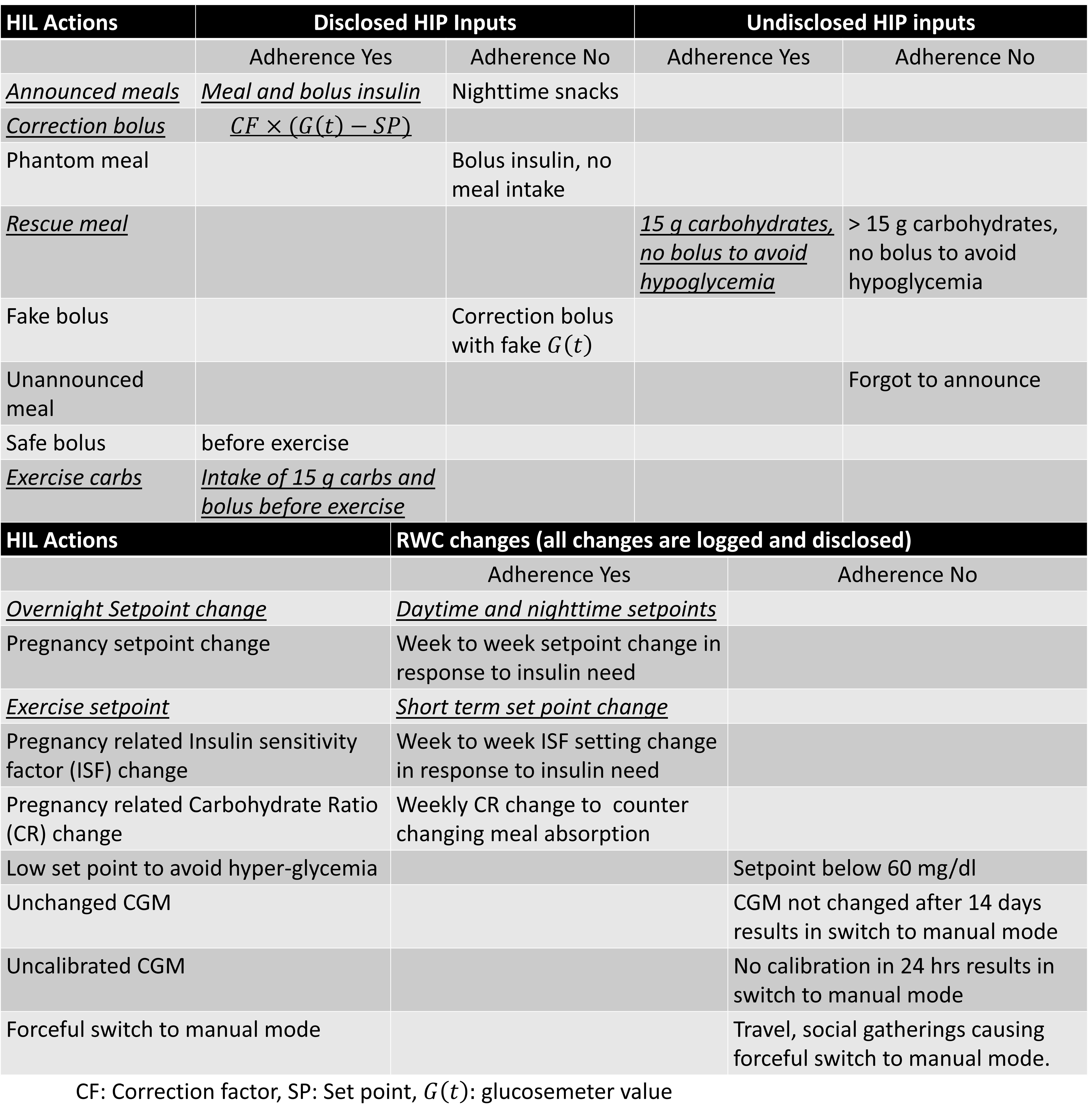}
\caption{Categorization of HIL actions.}
\label{fig:HumanCat}
\end{figure}

\noindent In \textbf{control affine systems}, the response of the plant to control inputs is modeled as a linear combination of the unperturbed continuous time state evolution of the plant, $f(X)$ and control input effect, $g(X)$ for the input $u$ (Eqn \ref{eqn:Plant}). For T1D the Bergman minimal model (BMM)~\cite{welch1990minimal} expresses $f(X)$ and $g(X)$ in the form of a set of nonlinear differential equations. The control inputs or actions $u=\pi(X)$ is a function of the sensed state variables (optimization of objective as a function of CGM and insulin). Human inputs are external inputs $U_{ex}$ in addition to the control inputs $u$ (Fig. \ref{fig:HumanCat}). 

\noindent{\bf HIL as a controller:} Human users play an integral part in AID system usage (ControlIQ~\cite{breton2021one}), where they can provide spontaneous inputs (such as meal) to the HIP or take part in critical hazard mitigation (rescue meal to mitigate hypoglycemia). The actions can be categorized into (Fig. \ref{fig:HumanCat}): a) disclosed or undisclosed HIP inputs, b) adherence to clinician guidance, and c) changes to RWC configuration. The underlined and italicized text denotes the profile of the average user for which the RWC is safety certified. There are several HIL actions that may adhere to clinician guidelines but are not commensurate with the average user profile, and are not certified safe.

The HIL-HIP \textbf{AS} (Fig. \ref{fig:HIL-HIP}) consists of AI engines that learn from human interaction and give personalization plans to the RWC, which is a temporally aligned finite sequence of control tasks interleaved with external inputs ($x^g_i(t_1) x^g_i(t_2)$ $ u^1_{ex} x^g_i(t_3)$ $ u^2_{ex}$ $ x^g_i(t_4) \ldots $) driven by actions in Fig. \ref{fig:HumanCat}. The \textbf{safety} is defined on subsets of $\mathcal{X}$, using Signal Temporal Logic (STL) $\phi$\cite{donze2010robust}. Eventual safety indicates that $\exists \tau \in \mathcal{R}: \phi |= True \forall t > \tau$, i.e. $\phi$ will be satisfied and remain true after some time $\tau$.



\noindent{\bf Safety certificate} for an RWC $\pi(X)$ is the existence of a given subset $C\subset\mathcal{X}$, with forward invariance ($\mathbb{FI}$) property~\cite{dawson2022safe}, which states that if the initial state of the AS is in $C$ then the closed loop system dynamics in Eqn \ref{eqn:Plant} keeps the initial state within $C$ in absence of any external perturbation. A set C has $\mathbb{FI}$ property if there exists a control lyapunov function $V(X), \forall X \in C$ such that $\forall X \in C$, $V(X) > 0$, $V(X^g) = 0$, for the set point $X^g$, and Eqn \ref{eqn:LPV} holds true. 
\begin{equation}
    \label{eqn:LPV}
    \footnotesize
    \forall X \in C, \exists \lambda >0 : L_f V(X) + L_g V(X) \pi(X) + \lambda V(X) < 0, 
\end{equation}
where $L_f$ and $L_g$ are lie derivatives of $V(X)$ along the direction of $f(X)$ and $g(X)$, respectively. 

\noindent{\bf Operational Safety:} An AS has operational safety if the safety STL satisfied by the AS model is also satisfied in real world deployment. 

  \subsection{Formal Problem Statement}
Given:\begin{itemize}
 \item a RWC, $\pi_{nom}(X)$
 \item a set $C \in \mathcal{X}$ such that $\forall X \in C$ at $t = 0$, the trajectory $X(t)\in C | \dot{X(t)} = f(X(t))+g(X(t))\pi_{nom}(X(t)), t > 0$, 
 \item a safety tuning probability (STP) value $p$
 \item a set of personalized inputs $U_{ex}=\{u_{ex}\}$.
\end{itemize}

\noindent   Find: a RWC $\pi_{NN}(X)$ such that $\forall X \in C$, for any $t > 0$  
   \begin{equation}
   \footnotesize
        P(X(t)\in C) > p  | \dot{X(t)} = f(X(t))+g(X(t))(\pi_{NN}(X(t))+u_{ex}),    
   \end{equation}

   \noindent{\bf Significance of HIL-HIP architecture:} In traditional control system design, once an AS is designed for average user there is no way to evaluate whether the deployed system will be safe for the real world user. However, in the HIL-HIP architecture the safe design is parameterized by STP $p$, which implies that the design is safe for human actions whose probability of occurrence is $p$ based on a human action model. Large value of $p$ indicates that the AS is safe for most common actions, whereas a smaller value of $p$ requires safe operation for even unusual actions. CLBF is easier to find with larger $p$ (p = 0.95 in the AID example in Section \ref{sec:Eval}) however, with smaller $p$ CLBF may not be found and hence the AS may not be certified safe. STP can be used by real user and their advisors to determine if their action model is compliant with the safety certificate. Traditional control system designs do not provide such safe personlization during operation.  
    \begin{figure}
\includegraphics[width=0.9\columnwidth]{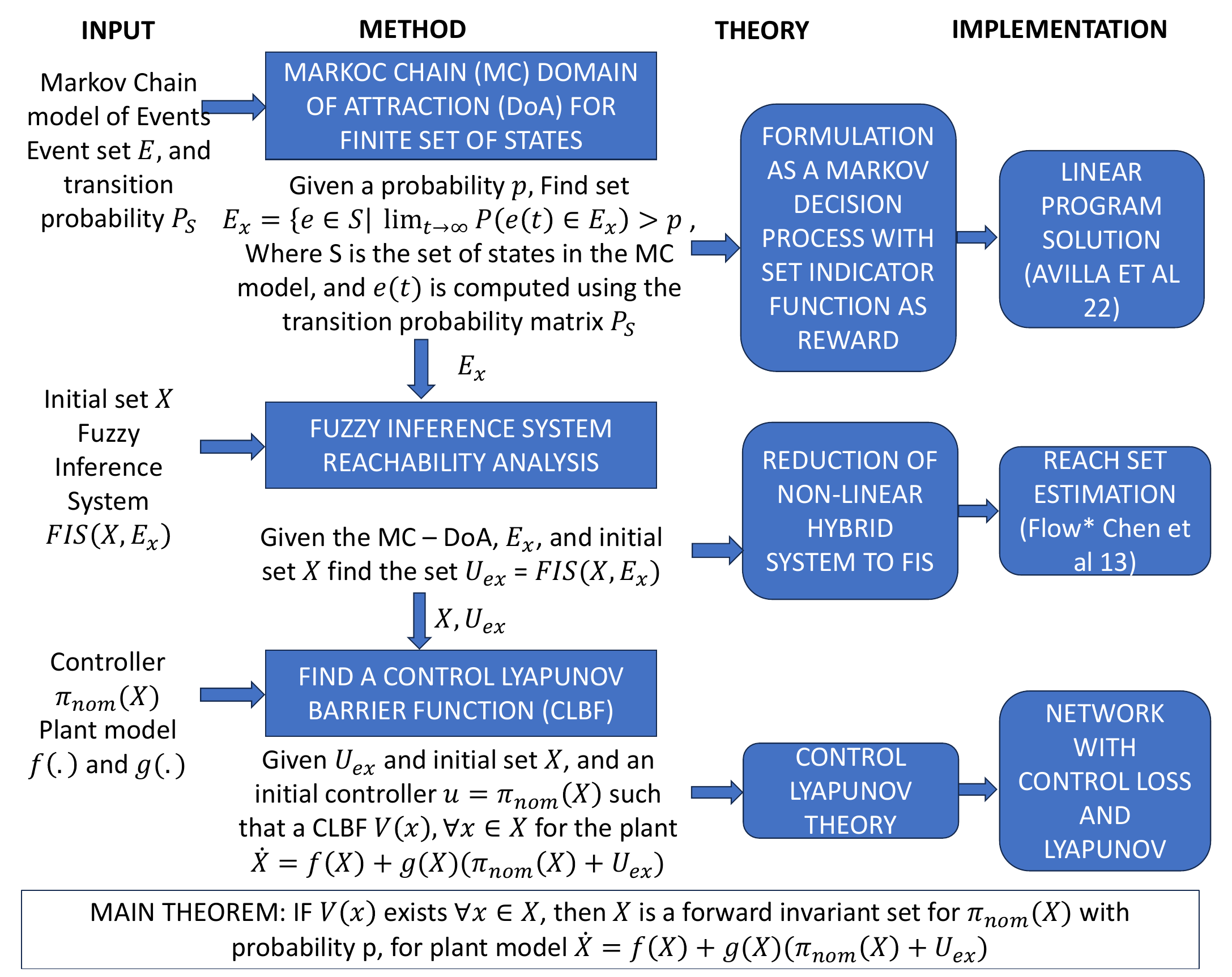}
\caption{Solution approach}
\label{fig:POS}
\end{figure}
\section{Solution and proof of safety}
To address safety of AS under personalized human actions, the AS is modeled as a HIL-HIP unified system, with three way interactions (Fig. \ref{fig:HIL-HIP}) between the HIL which is modeled as a combination of Markov Chain (MC) for spontaneous events and fuzzy inference system (FIS) for human responses to the events, RWC and the HIP. The HIL controller receives information from the RWC through data analysis/visualization applications and the natural biological feedback mechanism from the HIP. Based on advice from external advisory agents (clinicians), the HIL controller decides on: a) inputs to the HIP, such as meal or extra bolus insulin, and b) inputs to the RWC, such as settings change. Our method to derive safe HIL-HIP architecture consists of the following steps (Fig. \ref{fig:POS}):

    \noindent{\bf Step 1:} Find the domain of attraction (DoA) of MC model $E_x$ with minimum probability $p$. Starting from an initial set of states $E_I$ the DoA is the set of MC states $E_x$ that will occur at some point of time with atleast $p$ probability of occurrence. The MC extends a Markov Decision Process (MDP) with reward function same as the indicator function for the set $E_x$~\cite{Avila22On}. Solution of a linear program for value function maximization gives the DoA~\cite{Avila22On}.\\
    \noindent{\bf Step 2:} For the DoA $E_x$, find the reach set $U_{ex}$ of the FIS model for an initial set $X$ of states. We show that a hybrid system can be reduced to a FIS model. The reach set estimation method of the hybrid system model of the FIS gives an over-approximation of the reach set $U_{ex}$.\\
    \noindent{\bf Step 3:} For the $U_{ex}$, search a CLBF $V(X)$ using the neural architecture with control loss and lypunov loss to derive $\pi_{NN}$ with the plant model of Eqn. \ref{eqn:FG}.

\begin{figure}
\includegraphics[width=0.9\columnwidth]{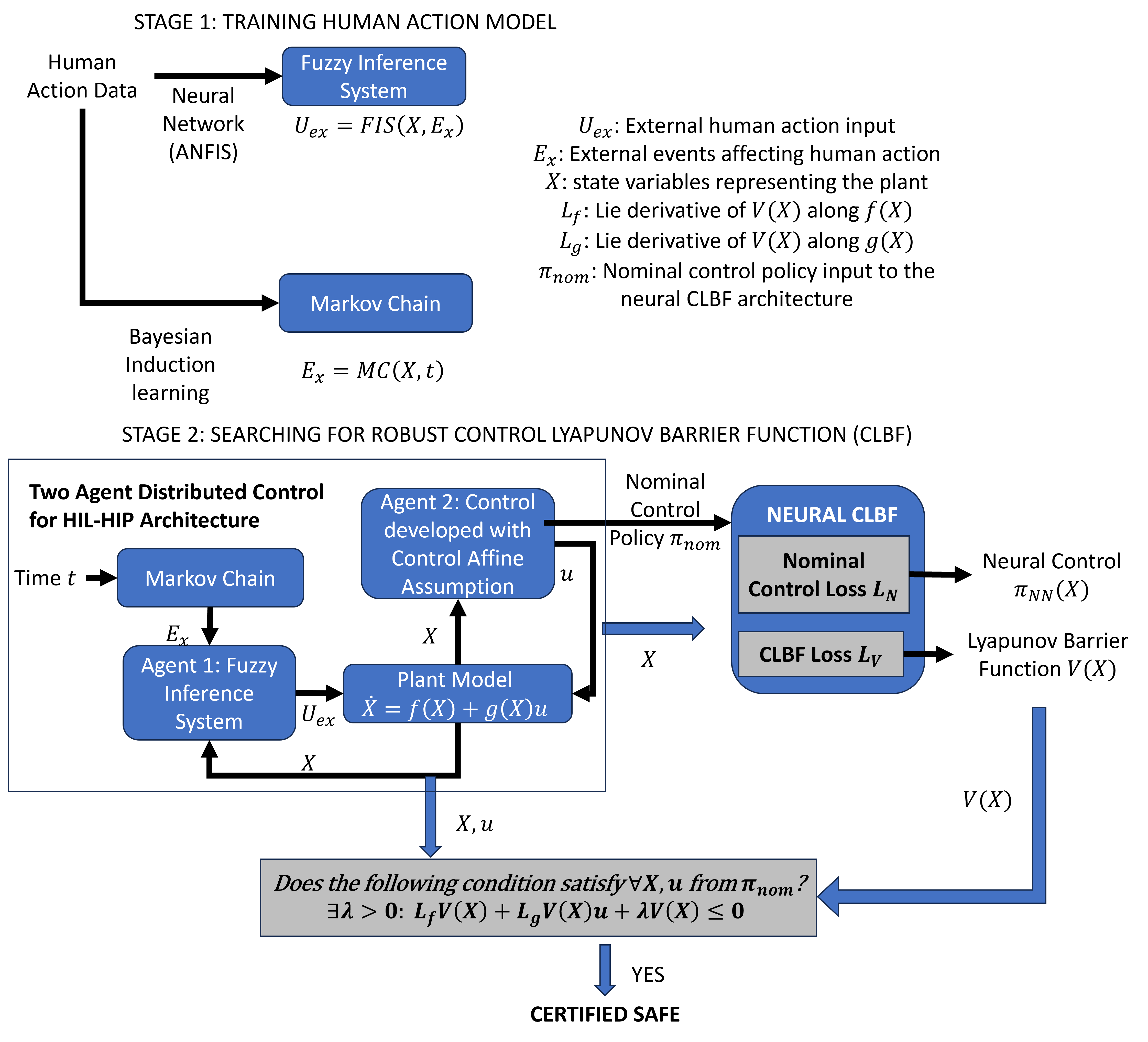}
\caption{Solution method for deriving safety certificate for HIL=HIP systems under the learned human aciton model}
\label{fig:TwoStage}
\end{figure}

To develop the HIL-HIP system, data driven learning is performed in two stages (Fig. \ref{fig:TwoStage}). In the first stage, the MC and FIS are learned which decide on human action. This human action is then fed to the second agent, which is the RWC learned using a neural architecture. 

\subsection{Modeling external events with Markov Chain}
The MC model is used to capture the spontaneous events that occur during long term usage of the AS. MC model states denote the unique events that are triggered due to the day to day activities of the HIP. Each unique event is parsed from the usage data of the AS. The set of unique events is denoted by $\mathbf{S}$. The transition from state $s_i\in \mathbf{S}$ to $s_j$ is tabulated from the usage data. Transition probabilities are computed using a conditional probability computation method and Bayes theorem. For each event $s_i$, we search the set of other events $\mathbf{S}_i = \{s_j\}$ such that $s_j$ is the next state after $s_i$. For each $s_j$ we count the number of times $n_{ij}$ that event $s_j$ occurred after event $s_i$. The transition probability $P_S(i,j)$, for transiting from event $s_i$ to $s_j$ is computed as $P_S(i,j) = \frac{n_{ij}}{\sum_{\forall s_j \in \mathbf{S}_i}{n_{ij}}}$. The MC model is thus represented by the state set $\mathbf{S}$ and transition probability matrix $P_S(.,.)$. 
\subsection{Finding DoA of MC with minimum probability $p$}
A MDP with the state space as $S$ and the reward function as the indicator function $\indicator{E_x}$ is considered, where $E_x$ is the DoA. The value function of the MDP as Eqn \ref{eqn:MDPV}, 
\begin{equation}
    \label{eqn:MDPV}
    \footnotesize
    v^{pol}(e) := \limsup_{M \to \infty} \frac{1}{M}E^\pi_e\big [ \sum^{M-1}_{t=0}{\indicator{E_x}(S)}\big], 
\end{equation}
where $E^\pi_e$ is the expected value function given as $P_S(e_j|e)$, where $e_j$ is any next state and $v^{pol}(e)$ denotes the value function for a given policy $pol$ in state $e$. 

\begin{lemma}
\label{lem:Lem1}
    The set of states explored by the optimal policy with value function $v^*(e) > p$, gives the reach set of the MC $(S,P_S)$ starting from state $e$, where $v^*$ is given by Eqn \ref{eqn:MDPV}.
\end{lemma}
The lemma is derived from Theorem III.4 in \cite{Avila22On}. The DoA $E_x$ of the MC $(S,P_S)$ with minimum probability $p$ is obtained by solving the linear program ReaP in~\cite{Avila22On}.  

\noindent{\bf Computational complexity:} The solution to the linear program has a complexity of $O(N^3 log(N/\delta))$, where $\delta$ is the error tolerance~\cite{Cohen21Solving}. 

\subsection{Learning a FIS from data}

The human action database $\mathcal{D} = \{E_x,X(t),$ $U_{ex}\}$ consists of traces $X(t)$ of the state variables $X$ over time $t$, external events $E_x$ that either affect the AS controller configuration or affect the HIP component, and the human action, $U_{ex}$, taken by the human in response to observation of the state variables and external events. We model the human action as a function of $X,E_x$ using techniques such as FIS. The membership functions and fuzzy rules is learned from data using techniques such as adaptive network-based fuzzy inference system (ANFIS)~\cite{salleh2017adaptive}. The output is human action model $U_{ex} = FIS(X,E_x)$.   
\subsection{Finding reach set for FIS}
\begin{lemma}
\label{lem:Lem2}
    For every FIS output $u_{ex}$ for a given $X$ and event $E_x$, there exists an execution of a rectangular hybrid system $(Q,V,\mathcal{F}:2^V\to \mathcal{R}^{|V|},Inv )$ that provides the final output $u_{ex}$ of one of its continuous states $v\in V$.
\end{lemma}
We prove this lemma by construction. The variable set $V$ of the hybrid system is the same as the state vector $X$ of the FIS. A rule in the ANFIS is of the form ``IF $x_1(k) \in A_{11}$ $ \wedge \ldots \wedge $ $x_N(k) \in A_{1N}$ THEN $u_{ex} \in B_1$'', where $A_{i,j}$ is the membership function of the $i^{th}$ rule for the $j^{th}$ element of the state vector and $B_i$ is the membership function of the output for the $i^{th}$ rule. Each rule $R_i$ learned by ANFIS is modeled as a state of the hybrid system in the set $Q$. The flow equation $f\in \mathcal{F}$ of each state $R_i$ is $defuzzify(min_{j = 1 \to N}{(A_{i,j})})$. Defuzzification is done using the centroid mechanism. Each $A_{i,j}$ is a nonlinear sigmoid function resulting in a nonlinear continuous flow function on the power set of variables $V=X$ to the $|V|=N$ dimensional real space. The state transition condition $Inv$ in the hybrid system is instantaneous and occurs by default resulting in a rectangular timed automata. By construction, an execution of this non-linear hybrid system follows the exact computational steps taken by FIS to arrive at an output. Hence, the reach set of the hybrid system provides the reach set $U_{ex}$ of the FIS. In this paper, we use the $Flow*$~\cite{chen2013flow} technique to obtain reach set of the derived hybrid system, which gives guaranteed over-approximation of $U_{ex}$. 

\noindent{\bf Computational Complexity:} The worst case computational complexity is $O(K N^2)$, where $K$ is the number of time steps ahead for which the reach set is computed~\cite{chen2013flow}.
\subsection{Safety certificate generation method}

The HIL-HIP controller is the integration between the human action controller ($U_{ex}=FIS(X,E_x)$, where $E_x = MC(t)$) and the AS controller $\pi(X)$. The total output of the HIL-HIP system $\pi_{nom}$ is $u = \pi_{nom}(X) = U_{ex} \bigcap \pi(X)$. For this controller, given a subset $c \subset \mathcal{X}$, we want to ensure the $\mathbb{FI}$ property. This is done by showing the existence of a CLBF $V(X)$ defined for any $X \in C$. We use neural architectures with modified loss functions discussed in~\cite{dawson2022safe} for this purpose. The main idea is to use the neural architecture as aproximators for: a) CLBF $V(X)$, and b) a neural controller $\pi_{NN}$ that satisfies the Lyapunov condition for stability and safety (Eqn. \ref{eqn:LPV}). The loss function consists of two parts: 

\noindent{\bf a) CLBF loss}, which ensures that the CLBF estimate by the neural structure $V(x)$ satisfies the relaxed condition of $V(x)<\epsilon > 0$, a small quantity.  

\noindent{\bf b) control loss}, which captures the difference in control actions by the neural controller $\pi_{NN}$ and the RWC $\pi_{nom}$. 

The neural structure was trained with a set of state variables and control actions from the RWC $\pi_{nom}$. The output of the training phase is: a) decision whether a CLBF exists or not, and b) if a CLBF exists then the trained neural architecture that gives both $\pi_{NN}$ and $V(X), \forall X \in C$, According to the CLBF theory~\cite{dawson2022safe}, if a $V(X)$ exists, then $\pi_{NN}$ is one of potentially many controllers (denoted by the set of controllers $\mathcal{K}(X)$) that are safe and the neural structure that gives $V(X)$ and the corresponding forward invariant set $C$ is a safety certificate. To ascertain whether $\pi_{nom} \in \mathcal{K}(X)$ we evaluate the CLBF condition in Eqn \ref{eqn:LPV} with $\pi_{nom}$ as the RWC.

\setcounter{theorem}{0}
\begin{theorem}
\label{th:Theo1}
If $V(X)$ exists $\forall X \in C$, then $C$ is a forward invariant set for $\pi_{NN}(X)$ with probability $p$, for the plant model in Eqn \ref{eqn:FG} and if condition in Eqn \ref{eqn:LPV} satisfies, then $C$ is also a forward invariant set for $\pi_{nom}(X)$ with probability $p$.
\end{theorem}
Lemma \ref{lem:Lem1} provides event set with occurrence probability $>p$. Lemma \ref{lem:Lem2} shows that $Flow*$ reachability analysis will provide a $U_{ex}$ that encompasses all FIS outputs that are $p$ probable due to the overapproximation property. Existence of the Lyapunov function through the neural structure guarantees that $\pi_{NN}$ will result in a safe plant when combined with $U_{ex}$. Hence, set $C$ will be forward invariant for $\pi_{NN}$. 

\noindent{\bf Computational complexity:} Discussed in Section \ref{sec:NN}
\section{Evaluation and Results}
\label{sec:Eval}
\subsection{Data Description}
We have accessed data from n = 20 patients with T1D for usage of the Tandem control IQ AID system for 22 weeks each (IRB information available). The patients were administered hydrocortisone dose of 40 mg, 20 mg, and 20 mg at 8 am, noon, and 2 pm on two supervised study days at study site.
\subsection{Safety violation in control affine assumption}
\begin{figure}
\centering
\includegraphics[width=\columnwidth,trim=0 50 0 0]{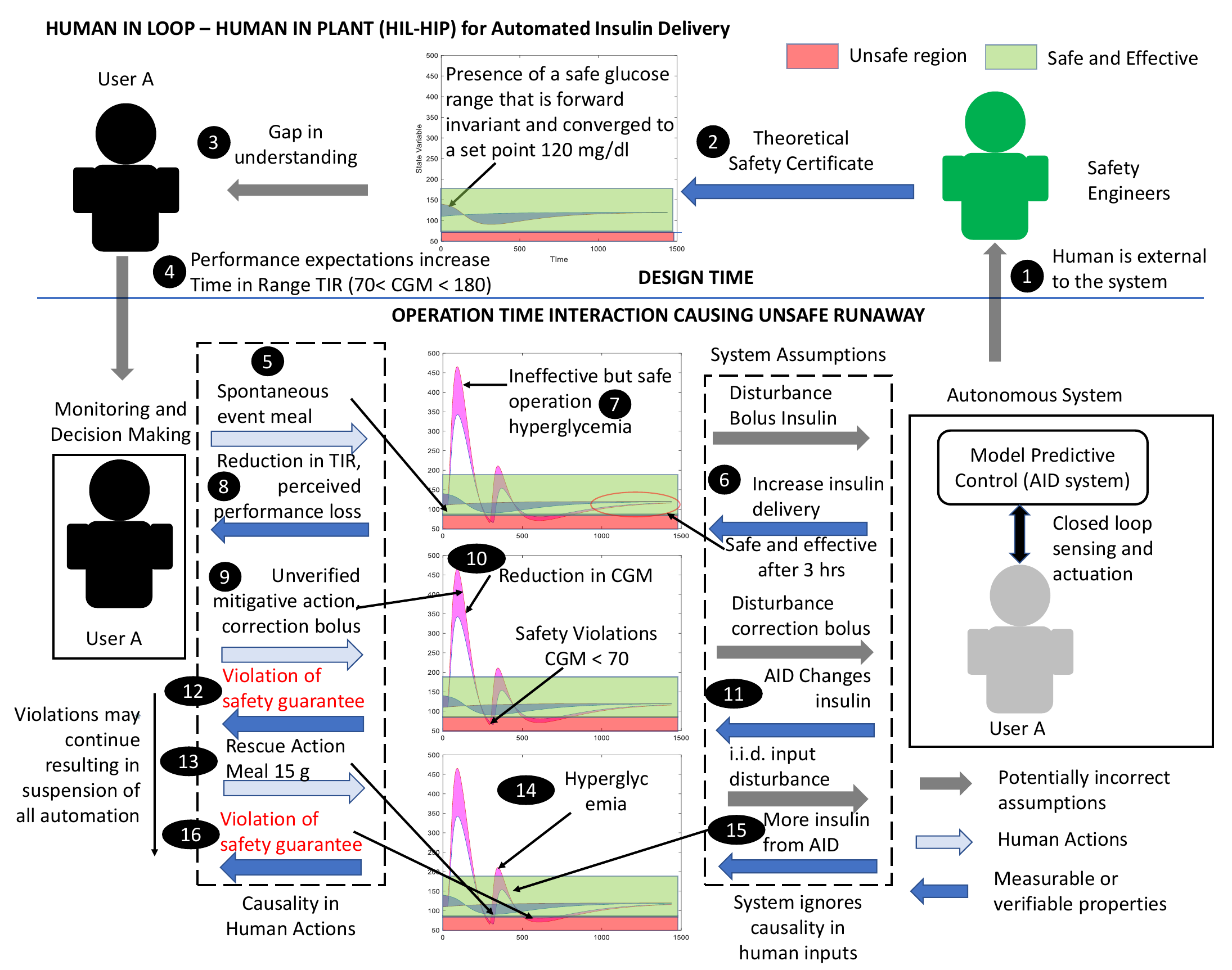}
\caption{Safety violations occur if human inputs are considered as external disturbances in AID systems for T1D. }
\label{fig:newSysMod}
\end{figure}
The illustrations in Fig. \ref{fig:newSysMod} is for an AID system, developed using the nonlinear optimal control theory discussed in Dawson et al~\cite{dawson2022safe}. Data from FDA approved T1D simulator is used to train a neural network with the CLBF loss function (code available in \texttt{MIT-REALM/neural\_clbf}). The multi-layer perceptron network was trained using 20,000 simulation data points to learn a CLBF. The set of initial glucose [110 mg/dl - 140 mg/dl] showed the $\mathbb{FI}$ property in absence of manual inputs, since the neural CLBF controller always keeps state trajectories within the initial set (shown by gray band in Fig. \ref{fig:newSysMod}). CLBF safety certificate generation mechanisms (Step 1 - 4 in Fig. \ref{fig:newSysMod})~\cite{dawson2022safe} applied to AID system assumes the T1D patient as external with meal and correction bolus insulin as independent identically distributed (i.i.d) random disturbances. A safety certified AID has large glucose excursions due to meal intake. However, it will ``eventually" enter a subset of the state space that has the $\mathbb{FI}$ property and hence be safe (no hypoglycemia glucose $> 70$ mg/dL). For a meal input of 100 g at 30 mins with 2U of rapid acting insulin suggested by the bolus wizard~\cite{andersen2016optimum} using ISF and CR settings, the system ``eventually" (i.e. after 3 hrs) brings the glucose to 120 mg/dl (the setpoint), pink band in Fig. \ref{fig:newSysMod}. 
 
  According to the ADA recommendations and physicians advice, 2 hrs after meal CGM should be within safe range, however, the patient observed CGM to be above 300 mg/dl. The patient following clinician guidelines came up with a plan of using a correction bolus computed as: $CB = (300 - 120)/CF$, for a correction factor (CF) of 20 mg/U, resulting in $9 U$ of insulin bolus. This resulted in the trajectory that cause hypoglycemia (steps 5 -7, intersection of pink band with red region in Fig. \ref{fig:newSysMod}).

 However, these interventions do not have safety guarantees and hence can lead the system to unsafe states (hypo-glycemia, glucose $<$ 70 mg/dL in Steps 8 - 12 in Fig. \ref{fig:newSysMod}). An unsafe excursion (CGM < 70 mg/dl) prompts the human to take immediate rescue carbohydrate of 15 g which drives up the glucose but takes it above 180 mg/dL, when the user could decide on getting another correction bolus. This can continue in operation time resulting in interaction runaway scenarios. The AS operation must be suspended to fail-safe modes. This is seen in nearly all AID systems such as Tandem Control IQ \cite{breton2021one}. Utilizing the theory presented in the paper, we use the MPC Control IQ strategy as $\pi_{nom}$ and learn $\pi_{NN}$ to avoid hypoglycemia in presence of human inputs.
\subsection{FIS model accuracy}
\begin{figure}
\includegraphics[width=0.5\columnwidth,trim=0 50 0 0]{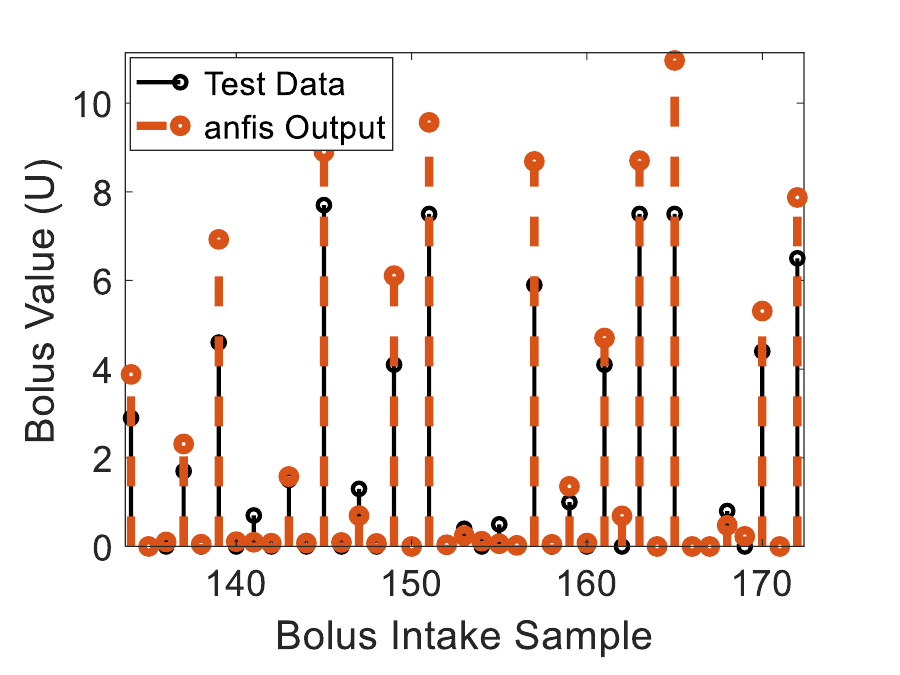}
\caption{Bolus prediction accuracy for ANFIS.}
\label{fig:Pred}
\end{figure}

We utilized an ANFIS~\cite{salleh2017adaptive} to predict individual external insulin bolus intake. There were an average of 522 ($\pm 15$) meals and meal boluses and 261 ($\pm 30$) correction bolus without meal. The ANFIS was designed with a $3 \times 1$ input vector consisting of $\{$ mean CGM, insulin on board computed using FIASP insulin action curve~\cite{grosman2021fast}, carbohydrate intake$\}$, and the output value of insulin bolus. With an 80-20 train-test split, the ANFIS achieved an RMSE of 0.45 in predicting insulin bolus (Fig. \ref{fig:Pred} shows prediction results). Execution time of the FIS model learning using the Matlab R2022 toolbox was 652 sec on an Intel Core i7 processor (single core). 

\subsection{MC modeling accuracy}
The MC had three states $\{$ Large, Medium, Small $\}$ indicating meal sizes. The T1D dataset was used to obtain three clusters of meal sizes for each individual. Utilizing 522 meal instances, the transition probabilities of the 3 state MC was learned. Monte Carlo simulation of the MC gave the distribution of each meal size, which matched with the distribution in real data $(p = 0.041)$. Execution time of MC learning was 0.267s.
\subsection{Performance of safety certified HIL-HIP system}
\label{sec:NN}
 \begin{figure}
\includegraphics[width=\columnwidth,trim=0 20 0 0]{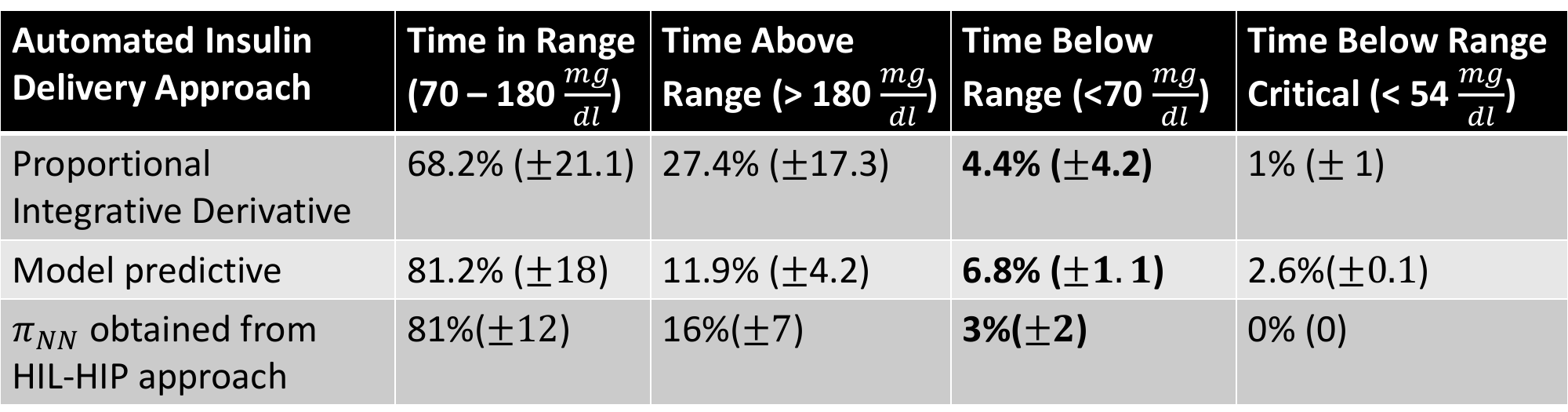}
\caption{Hypoglycemia reduction with HIL-HIP model.}
\label{fig:Compare}
\end{figure}

$\pi_{nom}$ was developed as a Model Predictive Control (MPC). A Bayesian meal prediction scheme utilizing the MC was integrated with the FIS to obtain the meal and correction bolus information. The integrated HIL-HIP system was simulated in closed loop with a Python implementation of Runge-Kutta solution of the BMM. The Neural CLBF architecture was then used with the MPC + Bayesian + FIS control as $\pi_{nom}$ which learned a CLBF and a new controller $\pi_{NN}$. The neural CLBF architecture is an MLP with 128 hidden layers. The input dimension is $3 \times 288$, where a single day CGM, Interstitial insulin and blood insulin was delivered as input. The sigmoid activation function was used in each neuron. The neural CLBF MLP was trained for 200 epochs which took 22 hrs on a 8 core Intel i7 CPU.  

We compare performance of $\pi_{NN}$ with regular MPC and Proportional Integrative and Derivative (PID) using 5 subjects in the T1D simulator~\cite{man2014uva} in Fig. \ref{fig:Compare}, which shows significant reduction in hypoglycemia for $\pi_{NN}$.
\section{Related Works}
Three broad classes of controller synthesis exist- {\bf a) Optimization approach:} For linear systems with eventual guarantees, a linear quadratic gaussian (LQG) optimal control strategy exists~\cite{karaman2008optimal}, which guarantees that a safety related STL will be satisfied. For non-linear systems with eventual guarantees, control Lyapunov (CLF) theory  exists~\cite{richards2018lyapunov}, which guarantees safety in absense of human inputs. 

\noindent{\bf b) Game theoretic approach:} The controller synthesis problem has been modeled as a two player game between the environment and the controller for safe HIL control~\cite{li2014synthesis}. These methods work well for 1D decision problems such as detection of safe switching time.

\noindent{\bf c) Reinforcement learning approach:} Safe RL is an emerging approach that models agents with a value function that has control objective as the reward and safety violation as the penalty function~\cite{garcia2015comprehensive}. Safe RL technique starts an initial safe MPC design that may not be effective, and for each control step evaluates the value function. If the value function is less than a threshold indicating heavy penalty, the safe RL defaults to the MPC strategy, else it uses the strategy obtained by maximizing the value function. This approach has been frequently used in robotics, however, the value function evaluation strategy does not involve human inputs.

To the best of our knowledge all attempts consider human inputs as external disturbances and as such may result in interaction runaway discussed in Fig. \ref{fig:newSysMod}. 
\section{Conclusions}
The paper presents extensions of safety certification theory to enable operational safety assurance of HIL-HIP AS. Departing from the traditional approach of assuming human inputs as external disturbances, this paper considers the human and AS as a co-operating unified system. The novel integration of Markov chains with fuzzy inference system to model human control and the neural control architecture to synthesize safe controller provides a mechanism for developing HIL-HIP AS as a unified system. This can provide early feedback on the safety of the AS operation so that mitigative actions can be taken proactively to avoid fatal accidents. We show the application of the new theory on AID controller synthesis for T1D, where HIL-HIP AID is shows to outperform commonly used AID systems such as MPC and PID with respect to safety and efficacy metrics.

\bibliographystyle{ACM-Reference-Format}
\bibliography{samples/ref,samples/NSFSATC}


\begin{thebibliography}{25}


\ifx \showCODEN    \undefined \def \showCODEN     #1{\unskip}     \fi
\ifx \showDOI      \undefined \def \showDOI       #1{#1}\fi
\ifx \showISBNx    \undefined \def \showISBNx     #1{\unskip}     \fi
\ifx \showISBNxiii \undefined \def \showISBNxiii  #1{\unskip}     \fi
\ifx \showISSN     \undefined \def \showISSN      #1{\unskip}     \fi
\ifx \showLCCN     \undefined \def \showLCCN      #1{\unskip}     \fi
\ifx \shownote     \undefined \def \shownote      #1{#1}          \fi
\ifx \showarticletitle \undefined \def \showarticletitle #1{#1}   \fi
\ifx \showURL      \undefined \def \showURL       {\relax}        \fi
\providecommand\bibfield[2]{#2}
\providecommand\bibinfo[2]{#2}
\providecommand\natexlab[1]{#1}
\providecommand\showeprint[2][]{arXiv:#2}

\bibitem[Andersen et~al\mbox{.}(2016)]%
        {andersen2016optimum}
\bibfield{author}{\bibinfo{person}{AJB Andersen}, \bibinfo{person}{Anne Ostenfeld}, \bibinfo{person}{Christian~Bressen Pipper}, \bibinfo{person}{Birthe~Susanne Olsen}, \bibinfo{person}{Anne~Marie Hertz}, \bibinfo{person}{Lene~K{\o}lle J{\o}rgensen}, \bibinfo{person}{Jette H{\o}gsmose}, {and} \bibinfo{person}{Jannet Svensson}.} \bibinfo{year}{2016}\natexlab{}.
\newblock \showarticletitle{Optimum bolus wizard settings in insulin pumps in children with Type 1 diabetes}.
\newblock \bibinfo{journal}{\emph{Diabetic Medicine}} \bibinfo{volume}{33}, \bibinfo{number}{10} (\bibinfo{year}{2016}), \bibinfo{pages}{1360--1365}.
\newblock


\bibitem[Banerjee and Gupta(2015)]%
        {Banerjee15TMCV2}
\bibfield{author}{\bibinfo{person}{Ayan Banerjee} {and} \bibinfo{person}{Sandeep~K.S. Gupta}.} \bibinfo{year}{2015}\natexlab{}.
\newblock \showarticletitle{Analysis of Smart Mobile Applications for Healthcare under Dynamic Context Changes}.
\newblock \bibinfo{journal}{\emph{IEEE Transactions on Mobile Computing}} \bibinfo{volume}{14}, \bibinfo{number}{5} (\bibinfo{year}{2015}), \bibinfo{pages}{904--919}.
\newblock
\urldef\tempurl%
\url{https://doi.org/10.1109/TMC.2014.2334606}
\showDOI{\tempurl}


\bibitem[Banerjee and Gupta(2012)]%
        {banerjee2012-your-mobility}
\bibfield{author}{\bibinfo{person}{Ayan Banerjee} {and} \bibinfo{person}{Sandeep K.~S. Gupta}.} \bibinfo{year}{2012}\natexlab{}.
\newblock \showarticletitle{Your mobility can be injurious to your health: Analyzing pervasive health monitoring systems under dynamic context changes}. In \bibinfo{booktitle}{\emph{Pervasive Computing and Communications (PerCom), 2012 IEEE International Conference on}}. \bibinfo{pages}{39 --47}.
\newblock
\urldef\tempurl%
\url{https://doi.org/10.1109/percom.2012.6199847}
\showDOI{\tempurl}


\bibitem[Banerjee et~al\mbox{.}(2023)]%
        {Banerjee23High}
\bibfield{author}{\bibinfo{person}{Ayan Banerjee}, \bibinfo{person}{Payal Kamboj}, \bibinfo{person}{Aranyak Maity}, \bibinfo{person}{Riya Salian}, {and} \bibinfo{person}{Sandeep Gupta}.} \bibinfo{year}{2023}\natexlab{}.
\newblock \showarticletitle{High Fidelity Fast Simulation of Human in the Loop Human in the Plant (HIL-HIP) Systems}. In \bibinfo{booktitle}{\emph{Proceedings of the Int'l ACM Conference on Modeling Analysis and Simulation of Wireless and Mobile Systems}} (Montreal, Quebec, Canada) \emph{(\bibinfo{series}{MSWiM '23})}. \bibinfo{publisher}{Association for Computing Machinery}, \bibinfo{address}{New York, NY, USA}, \bibinfo{pages}{199–203}.
\newblock
\showISBNx{9798400703669}
\urldef\tempurl%
\url{https://doi.org/10.1145/3616388.3617550}
\showDOI{\tempurl}


\bibitem[Banerjee et~al\mbox{.}(2024)]%
        {banerjee2024cpsllmlargelanguagemodel}
\bibfield{author}{\bibinfo{person}{Ayan Banerjee}, \bibinfo{person}{Aranyak Maity}, \bibinfo{person}{Payal Kamboj}, {and} \bibinfo{person}{Sandeep K.~S. Gupta}.} \bibinfo{year}{2024}\natexlab{}.
\newblock \bibinfo{title}{CPS-LLM: Large Language Model based Safe Usage Plan Generator for Human-in-the-Loop Human-in-the-Plant Cyber-Physical System}.
\newblock
\newblock
\showeprint[arxiv]{2405.11458}~[cs.AI]
\urldef\tempurl%
\url{https://arxiv.org/abs/2405.11458}
\showURL{%
\tempurl}
\newblock
\shownote{AI Planning for Cyber-Physical Systems – CAIPI’24 AAAI}.


\bibitem[Banerjee et~al\mbox{.}(2011)]%
        {banerjee2011ensuring}
\bibfield{author}{\bibinfo{person}{Ayan Banerjee}, \bibinfo{person}{Krishna~K Venkatasubramanian}, \bibinfo{person}{Tridib Mukherjee}, {and} \bibinfo{person}{Sandeep Kumar~S Gupta}.} \bibinfo{year}{2011}\natexlab{}.
\newblock \showarticletitle{Ensuring safety, security, and sustainability of mission-critical cyber--physical systems}.
\newblock \bibinfo{journal}{\emph{Proc. IEEE}} \bibinfo{volume}{100}, \bibinfo{number}{1} (\bibinfo{year}{2011}), \bibinfo{pages}{283--299}.
\newblock


\bibitem[Banerjee et~al\mbox{.}(2013)]%
        {banerjee2013using}
\bibfield{author}{\bibinfo{person}{Ayan Banerjee}, \bibinfo{person}{Yi Zhang}, \bibinfo{person}{Paul Jones}, {and} \bibinfo{person}{Sandeep Gupta}.} \bibinfo{year}{2013}\natexlab{}.
\newblock \showarticletitle{Using formal methods to improve home-use medical device safety}.
\newblock \bibinfo{journal}{\emph{Biomedical instrumentation \& technology}} \bibinfo{volume}{47}, \bibinfo{number}{s1} (\bibinfo{year}{2013}), \bibinfo{pages}{43--48}.
\newblock


\bibitem[Breton and Kovatchev(2021)]%
        {breton2021one}
\bibfield{author}{\bibinfo{person}{Marc~D Breton} {and} \bibinfo{person}{Boris~P Kovatchev}.} \bibinfo{year}{2021}\natexlab{}.
\newblock \showarticletitle{One year real-world use of the control-IQ advanced hybrid closed-loop technology}.
\newblock \bibinfo{journal}{\emph{Diabetes Technology \& Therapeutics}} \bibinfo{volume}{23}, \bibinfo{number}{9} (\bibinfo{year}{2021}), \bibinfo{pages}{601--608}.
\newblock


\bibitem[Chen et~al\mbox{.}(2013)]%
        {chen2013flow}
\bibfield{author}{\bibinfo{person}{Xin Chen}, \bibinfo{person}{Erika {\'A}brah{\'a}m}, {and} \bibinfo{person}{Sriram Sankaranarayanan}.} \bibinfo{year}{2013}\natexlab{}.
\newblock \showarticletitle{Flow*: An analyzer for non-linear hybrid systems}. In \bibinfo{booktitle}{\emph{Computer Aided Verification: 25th International Conference, CAV 2013, Saint Petersburg, Russia, July 13-19, 2013. Proceedings 25}}. Springer, \bibinfo{pages}{258--263}.
\newblock


\bibitem[Cohen et~al\mbox{.}(2021)]%
        {Cohen21Solving}
\bibfield{author}{\bibinfo{person}{Michael~B. Cohen}, \bibinfo{person}{Yin~Tat Lee}, {and} \bibinfo{person}{Zhao Song}.} \bibinfo{year}{2021}\natexlab{}.
\newblock \showarticletitle{Solving Linear Programs in the Current Matrix Multiplication Time}.
\newblock \bibinfo{journal}{\emph{J. ACM}} \bibinfo{volume}{68}, \bibinfo{number}{1}, Article \bibinfo{articleno}{3} (\bibinfo{date}{jan} \bibinfo{year}{2021}), \bibinfo{numpages}{39}~pages.
\newblock
\showISSN{0004-5411}
\urldef\tempurl%
\url{https://doi.org/10.1145/3424305}
\showDOI{\tempurl}


\bibitem[Dawson et~al\mbox{.}(2022)]%
        {dawson2022safe}
\bibfield{author}{\bibinfo{person}{Charles Dawson}, \bibinfo{person}{Sicun Gao}, {and} \bibinfo{person}{Chuchu Fan}.} \bibinfo{year}{2022}\natexlab{}.
\newblock \showarticletitle{Safe control with learned certificates: A survey of neural lyapunov, barrier, and contraction methods}.
\newblock \bibinfo{journal}{\emph{arXiv preprint arXiv:2202.11762}} (\bibinfo{year}{2022}).
\newblock


\bibitem[Donz{\'e} and Maler(2010)]%
        {donze2010robust}
\bibfield{author}{\bibinfo{person}{Alexandre Donz{\'e}} {and} \bibinfo{person}{Oded Maler}.} \bibinfo{year}{2010}\natexlab{}.
\newblock \showarticletitle{Robust satisfaction of temporal logic over real-valued signals}. In \bibinfo{booktitle}{\emph{Formal Modeling and Analysis of Timed Systems: 8th International Conference, FORMATS 2010, Klosterneuburg, Austria, September 8-10, 2010. Proceedings 8}}. Springer, \bibinfo{pages}{92--106}.
\newblock


\bibitem[Garc{\i}a and Fern{\'a}ndez(2015)]%
        {garcia2015comprehensive}
\bibfield{author}{\bibinfo{person}{Javier Garc{\i}a} {and} \bibinfo{person}{Fernando Fern{\'a}ndez}.} \bibinfo{year}{2015}\natexlab{}.
\newblock \showarticletitle{A comprehensive survey on safe reinforcement learning}.
\newblock \bibinfo{journal}{\emph{Journal of Machine Learning Research}} \bibinfo{volume}{16}, \bibinfo{number}{1} (\bibinfo{year}{2015}), \bibinfo{pages}{1437--1480}.
\newblock


\bibitem[Grosman et~al\mbox{.}(2021)]%
        {grosman2021fast}
\bibfield{author}{\bibinfo{person}{Benyamin Grosman}, \bibinfo{person}{Di Wu}, \bibinfo{person}{Neha Parikh}, \bibinfo{person}{Anirban Roy}, \bibinfo{person}{Gayane Voskanyan}, \bibinfo{person}{Natalie Kurtz}, \bibinfo{person}{Jeppe Sturis}, \bibinfo{person}{Ohad Cohen}, \bibinfo{person}{Magnus Ekelund}, {and} \bibinfo{person}{Robert Vigersky}.} \bibinfo{year}{2021}\natexlab{}.
\newblock \showarticletitle{Fast-acting insulin aspart (Fiasp{\textregistered}) improves glycemic outcomes when used with MiniMedTM 670G hybrid closed-loop system in simulated trials compared to NovoLog{\textregistered}}.
\newblock \bibinfo{journal}{\emph{Computer Methods and Programs in Biomedicine}}  \bibinfo{volume}{205} (\bibinfo{year}{2021}), \bibinfo{pages}{106087}.
\newblock


\bibitem[Karaman et~al\mbox{.}(2008)]%
        {karaman2008optimal}
\bibfield{author}{\bibinfo{person}{Sertac Karaman}, \bibinfo{person}{Ricardo~G Sanfelice}, {and} \bibinfo{person}{Emilio Frazzoli}.} \bibinfo{year}{2008}\natexlab{}.
\newblock \showarticletitle{Optimal control of mixed logical dynamical systems with linear temporal logic specifications}. In \bibinfo{booktitle}{\emph{2008 47th IEEE Conference on Decision and Control}}. IEEE, \bibinfo{pages}{2117--2122}.
\newblock


\bibitem[Lamrani et~al\mbox{.}(2018)]%
        {lamrani2018hymn}
\bibfield{author}{\bibinfo{person}{Imane Lamrani}, \bibinfo{person}{Ayan Banerjee}, {and} \bibinfo{person}{Sandeep~KS Gupta}.} \bibinfo{year}{2018}\natexlab{}.
\newblock \showarticletitle{HyMn: Mining linear hybrid automata from input output traces of cyber-physical systems}. In \bibinfo{booktitle}{\emph{2018 IEEE Industrial Cyber-Physical Systems (ICPS)}}. IEEE, \bibinfo{pages}{264--269}.
\newblock


\bibitem[Lamrani et~al\mbox{.}(2021a)]%
        {waise}
\bibfield{author}{\bibinfo{person}{Imane Lamrani}, \bibinfo{person}{Ayan Banerjee}, {and} \bibinfo{person}{Sandeep K.~S. Gupta}.} \bibinfo{year}{2021}\natexlab{a}.
\newblock \showarticletitle{Certification Game for the Safety Analysis of AI-Based CPS}. In \bibinfo{booktitle}{\emph{Computer Safety, Reliability, and Security. SAFECOMP 2021 Workshops}}, \bibfield{editor}{\bibinfo{person}{Ibrahim Habli}, \bibinfo{person}{Mark Sujan}, \bibinfo{person}{Simos Gerasimou}, \bibinfo{person}{Erwin Schoitsch}, {and} \bibinfo{person}{Friedemann Bitsch}} (Eds.). \bibinfo{publisher}{Springer International Publishing}, \bibinfo{address}{Cham}, \bibinfo{pages}{297--310}.
\newblock
\showISBNx{978-3-030-83906-2}


\bibitem[Lamrani et~al\mbox{.}(2021b)]%
        {Lamrani21OperationalV2}
\bibfield{author}{\bibinfo{person}{Imane Lamrani}, \bibinfo{person}{Ayan Banerjee}, {and} \bibinfo{person}{Sandeep K.~S. Gupta}.} \bibinfo{year}{2021}\natexlab{b}.
\newblock \showarticletitle{Operational Data-Driven Feedback for Safety Evaluation of Agent-Based Cyber–Physical Systems}.
\newblock \bibinfo{journal}{\emph{IEEE Transactions on Industrial Informatics}} \bibinfo{volume}{17}, \bibinfo{number}{5} (\bibinfo{year}{2021}), \bibinfo{pages}{3367--3378}.
\newblock
\urldef\tempurl%
\url{https://doi.org/10.1109/TII.2020.3009985}
\showDOI{\tempurl}


\bibitem[Li et~al\mbox{.}(2014)]%
        {li2014synthesis}
\bibfield{author}{\bibinfo{person}{Wenchao Li}, \bibinfo{person}{Dorsa Sadigh}, \bibinfo{person}{S~Shankar Sastry}, {and} \bibinfo{person}{Sanjit~A Seshia}.} \bibinfo{year}{2014}\natexlab{}.
\newblock \showarticletitle{Synthesis for human-in-the-loop control systems}. In \bibinfo{booktitle}{\emph{Tools and Algorithms for the Construction and Analysis of Systems: 20th International Conference, TACAS 2014, Held as Part of the European Joint Conferences on Theory and Practice of Software, ETAPS 2014, Grenoble, France, April 5-13, 2014. Proceedings 20}}. Springer, \bibinfo{pages}{470--484}.
\newblock


\bibitem[Man et~al\mbox{.}(2014)]%
        {man2014uva}
\bibfield{author}{\bibinfo{person}{Chiara~Dalla Man}, \bibinfo{person}{Francesco Micheletto}, \bibinfo{person}{Dayu Lv}, \bibinfo{person}{Marc Breton}, \bibinfo{person}{Boris Kovatchev}, {and} \bibinfo{person}{Claudio Cobelli}.} \bibinfo{year}{2014}\natexlab{}.
\newblock \showarticletitle{The UVA/PADOVA type 1 diabetes simulator: new features}.
\newblock \bibinfo{journal}{\emph{Journal of diabetes science and technology}} \bibinfo{volume}{8}, \bibinfo{number}{1} (\bibinfo{year}{2014}), \bibinfo{pages}{26--34}.
\newblock


\bibitem[Priyanka~Bagade(2017)]%
        {bagade2017validation}
\bibfield{author}{\bibinfo{person}{Sandeep K.S.~Gupta Priyanka~Bagade, Ayan~Banerjee}.} \bibinfo{year}{2017}\natexlab{}.
\newblock \showarticletitle{Validation, verification, and formal methods for cyber-physical systems}.
\newblock In \bibinfo{booktitle}{\emph{Cyber-Physical Systems}}. \bibinfo{publisher}{Elsevier}, \bibinfo{pages}{175--191}.
\newblock


\bibitem[Richards et~al\mbox{.}(2018)]%
        {richards2018lyapunov}
\bibfield{author}{\bibinfo{person}{Spencer~M Richards}, \bibinfo{person}{Felix Berkenkamp}, {and} \bibinfo{person}{Andreas Krause}.} \bibinfo{year}{2018}\natexlab{}.
\newblock \showarticletitle{The lyapunov neural network: Adaptive stability certification for safe learning of dynamical systems}. In \bibinfo{booktitle}{\emph{Conference on Robot Learning}}. PMLR, \bibinfo{pages}{466--476}.
\newblock


\bibitem[Salleh et~al\mbox{.}(2017)]%
        {salleh2017adaptive}
\bibfield{author}{\bibinfo{person}{Mohd Najib~Mohd Salleh}, \bibinfo{person}{Noureen Talpur}, {and} \bibinfo{person}{Kashif Hussain}.} \bibinfo{year}{2017}\natexlab{}.
\newblock \showarticletitle{Adaptive neuro-fuzzy inference system: Overview, strengths, limitations, and solutions}. In \bibinfo{booktitle}{\emph{Data Mining and Big Data: Second International Conference, DMBD 2017, Fukuoka, Japan, July 27--August 1, 2017, Proceedings 2}}. Springer, \bibinfo{pages}{527--535}.
\newblock


\bibitem[Welch et~al\mbox{.}(1990)]%
        {welch1990minimal}
\bibfield{author}{\bibinfo{person}{S Welch}, \bibinfo{person}{SSP Gebhart}, \bibinfo{person}{RN Bergman}, {and} \bibinfo{person}{LS Phillips}.} \bibinfo{year}{1990}\natexlab{}.
\newblock \showarticletitle{Minimal model analysis of intravenous glucose tolerance test-derived insulin sensitivity in diabetic subjects}.
\newblock \bibinfo{journal}{\emph{The Journal of Clinical Endocrinology \& Metabolism}} \bibinfo{volume}{71}, \bibinfo{number}{6} (\bibinfo{year}{1990}), \bibinfo{pages}{1508--1518}.
\newblock


\bibitem[Ávila and Junca(2022)]%
        {Avila22On}
\bibfield{author}{\bibinfo{person}{Daniel Ávila} {and} \bibinfo{person}{Mauricio Junca}.} \bibinfo{year}{2022}\natexlab{}.
\newblock \showarticletitle{On Reachability of Markov Chains: A Long-Run Average Approach}.
\newblock \bibinfo{journal}{\emph{IEEE Trans. Automat. Control}} \bibinfo{volume}{67}, \bibinfo{number}{4} (\bibinfo{year}{2022}), \bibinfo{pages}{1996--2003}.
\newblock
\urldef\tempurl%
\url{https://doi.org/10.1109/TAC.2021.3071334}
\showDOI{\tempurl}


\end{thebibliography}

\end{document}